\documentclass[aps,11pt,onecolumn,preprintnumbers,amsmath,amssymb]{revtex4}

\usepackage[english]{babel}
\usepackage{graphicx}
\usepackage{natbib}
\usepackage{color}

\makeatletter
\@ifundefined{textcolor}{}
{%
 \definecolor{BLACK}{gray}{0}
 \definecolor{WHITE}{gray}{1}
 \definecolor{RED}{rgb}{1,0,0}
 \definecolor{GREEN}{rgb}{0,1,0}
 \definecolor{BLUE}{rgb}{0,0,1}
 \definecolor{CYAN}{cmyk}{1,0,0,0}
 \definecolor{MAGENTA}{cmyk}{0,1,0,0}
 \definecolor{YELLOW}{cmyk}{0,0,1,0}
 }

\begin{document}
\title{Electrically tunable artificial gauge potential for polaritons}
\author{Hyang-Tag Lim$^*$}
\author{Emre Togan$^*$}
\author{Martin Kroner}
\author{Javier Miguel-Sanchez}
\author{Atac Imamo\u{g}lu}
\affiliation{Institute of Quantum Electronics, ETH Zurich, CH-8093
Zurich, Switzerland.\\
$^*$These authors contributed equally to this work.}

\maketitle

\textbf{Neutral particles subject to artificial gauge potentials can behave as charged particles in magnetic fields. This fascinating premise has led to demonstrations of one-way waveguides, topologically protected edge states and Landau levels for photons. In ultracold neutral atoms effective gauge fields have allowed the  emulation of matter under strong magnetic fields leading to realization of Harper-Hofstadter and Haldane models. Here we show that application of perpendicular electric and magnetic fields effects a tuneable artificial gauge potential for two-dimensional microcavity exciton polaritons. For verification, we perform interferometric measurement of the associated phase accumulated during coherent polariton transport. Since the gauge potential originates from the magnetoelectric Stark effect, it can be realized for photons strongly coupled to excitations in any polarizable medium. Together with strong polariton-polariton interactions and engineered polariton lattices, artificial gauge fields could play a key role in investigation of non-equilibrium dynamics of strongly correlated photons.
}

Synthesis of artificial gauge fields for photons have been demonstrated in a number of different optical systems. In most cases, the implementation is achieved through the design of the optical system~\cite{Rechtsman2012, Rechtsman2013, Hafezi2013a,Schine2016}, leaving little or no room for fast control of the magnitude of the effected gauge field after sample fabrication is completed. For many applications on the other hand, it is essential to be able to tune or adjust the strength of the gauge field during
the experiment~\cite{Aidelsburger2013,Jotzu2014,Dalibard2011}; this is particularly the case for nanophotonic structures~\cite{Jacqmin2014} where fast local control of the gauge field strength may open up new possibilities for investigation of many-body physics of light~\cite{Carusotto2013}. The realization we describe here has the potential to fulfill this premise since the strength and direction of the effected gauge potential is controlled electrically. Moreover, since our scheme relies on the magnetoelectric Stark effect, time reversal symmetry~\cite{Wang2009,Fang2012,Tzuang2014} of the optical excitations is broken.

Cavity-polaritons are hybrid light-matter quasi-particles arising from non-perturbative coupling between quantum well (QW) excitons and cavity photons. A magnetic field $B_z$ applied along the growth direction influences polaritons through their excitonic nature and leads to a diamagnetic shift, Zeeman splitting of circularly polarized modes and enhancement of exciton-photon coupling strength~\cite{pietka2015}.  For excitons with a non-zero momentum, the applied $B_z$ also induces an  electric dipole moment~\cite{kallin1984, paquet1985, lozovik2002, butov2001}. In a classical picture, this dipole moment is due to the Lorentz force that creates an effective electric field $E_{{\rm eff}}$~\cite{Thomas1961} that is proportional to $\vec{k} \times \vec{B}$.  For excitons with small momentum and polarizability $\alpha$, this $E_{{\rm eff}}$ causes an induced dipole moment ($\vec{d} \propto \alpha \vec{k} \times \vec{B}$).  An additional external electric field $\vec E_{{\rm ext}}$ in the  QW plane then alters the dispersion of excitons as it leads to energy changes proportional to $k$: $- \vec{d} \cdot \vec E_{{\rm ext}} \propto -\alpha \left( \vec{k} \times \vec{B} \right) \cdot \vec E_{{\rm ext}}$, so that the energy minimum is no longer at $|\vec{k}| = 0$ as would be the case for a free particle with an effective mass but at finite $\vec{k}$. This simple modification of the dispersion relation is equivalent to an effective gauge potential $A_{{\rm eff}}$ for excitons; due to their partly excitonic character, the hamiltonian describing the dynamics of polaritons also contains a term proportional to $A_{{\rm eff}}$.

We first demonstrate the presence of an effective electric field whose magnitude and direction depends on the direction of propagation for polaritons subjected to an out-of-plane magnetic field. The cavity polariton sample we use is illustrated in Figure ~\ref{ideaFigure}a. Our demonstration relies on the fact that it is possible to excite polaritons with a well defined in-plane wavevector $\vec{k}$ by appropriately choosing the angle and energy of the  excitation beam. As illustrated   in Figure~\ref{ideaFigure}a we choose $\vec{k} = k_y \hat{y}$. Two metal gates deposited $30~\mu$m apart  allow us to apply an electric field in the $x$ direction such that for polaritons propagating with a wavevector along the $y$ direction,  $\vec E_{{\rm ext}}$ will add or cancel $E_{{\rm eff}}$ due to the Lorentz force. By recording changes in the transition energy of polaritons as a function of $E_{{\rm ext}}$ we can determine the strength of $E_{{\rm eff}}$. In our experiments, we only address the lower polariton branch.

Changes in the reflected intensity of a laser beam probing polaritons at $B_z = 0$ T with $k_y^+ = 2.7~\mu$m$^{-1}$ as a function of $E_{{\rm ext}}$ is shown in Figure~\ref{ideaFigure}d. For each $E_{{\rm ext}}$ the reflected intensity shows a dip at an energy corresponding to the polariton resonance (Figure~\ref{ideaFigure}e). The spectral center of the dip shifts to lower energies with $E_{{\rm ext}}$ in a way that is  well described by a second order polynomial. The expected behaviour for a neutral polarizable quasi-particle (i.e. exciton or polariton) that is subject to $E_{{\rm ext}}$, would be to have a dc-Stark shift proportional to $-\alpha \left |E_{{\rm ext}} \right |^2$~\cite{miller1985}; here, the coefficient $\alpha$ is the quasi-particle polarizability. Therefore the electric field that yields the  maximum lower polariton energy identifies $E_{{\rm ext}} = \tilde E_{{\rm ext}}$ that exactly cancels any internal or effective electric fields $E_{{\rm eff}}$ ($\tilde E_{{\rm ext}}=-E_{{\rm eff}}$). To quantify $E_{{\rm eff}}$ due to the Lorentz force described above,  we extract the difference of $\tilde E_{{\rm ext}}$ at a fixed $B_z$  for polaritons excited with $k_y^+ = 2.7~\mu$m$^{-1}$ and $k_y^- = -2.9~\mu$m$^{-1}$. This approach also allows us to exclude influences of built in electric fields. At $B_z = 0$ T we find that the difference of $V_G$ values that correspond to the maximum energy with $k_y^\pm$ excitation is 0.02 V, indicating that $E_{{\rm eff}}$ is negligible at 0 T (Figure~\ref{ideaFigure}f).

\begin{figure}
\includegraphics[width = \textwidth]{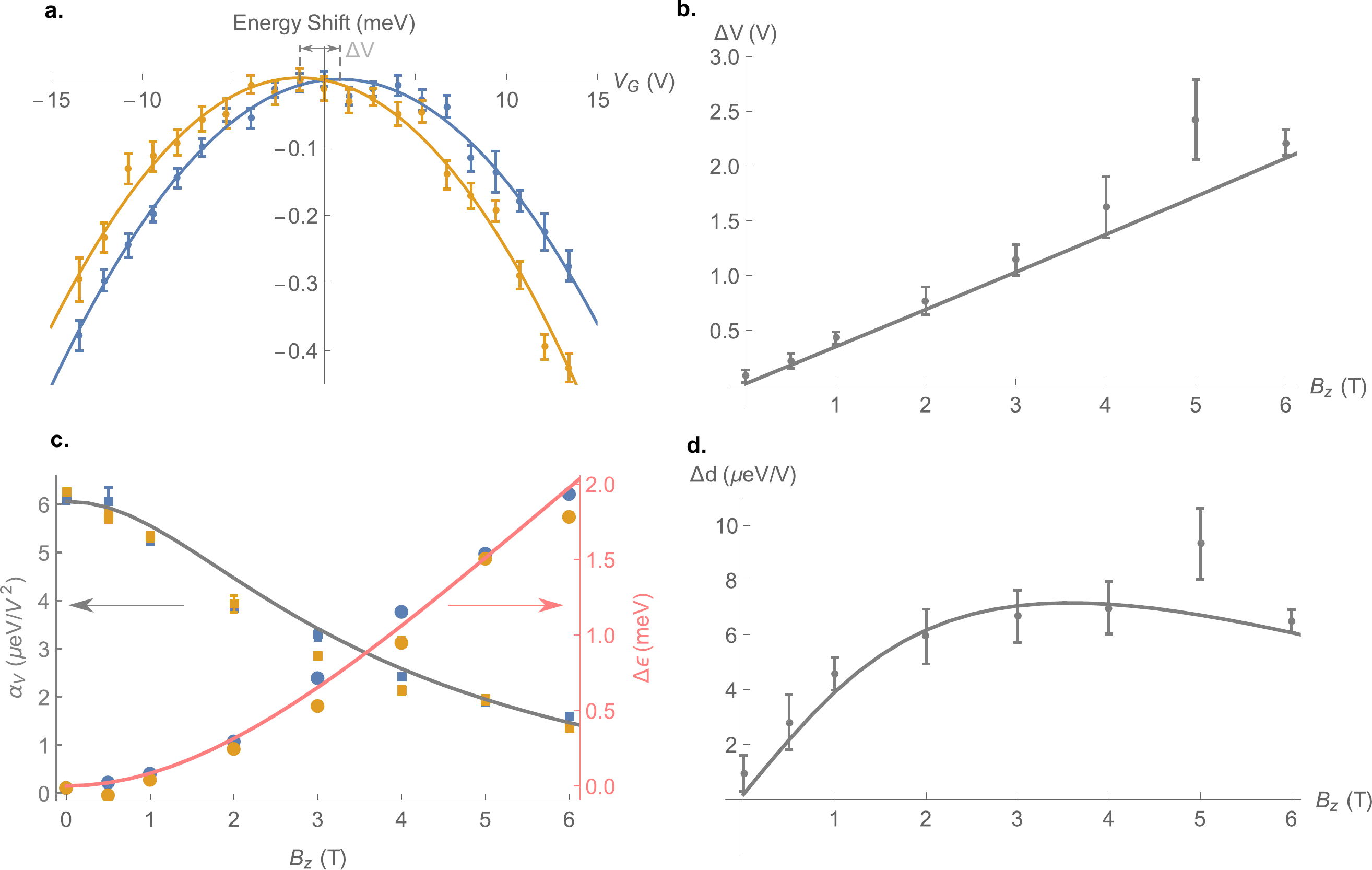}
\caption{\textbf{Magnetic field dependence.}
\textbf{a.} Shift of polariton energies with applied voltage for polaritons excited with  $k_y^+$ (yellow) and $k_y^- $ (blue) at $B_z$ = 5 T; solid lines are fits to second order polynomials ($-d_V V_G -\alpha_V V_G^2$, Blue: $d_V= 5.2 \times 10^{-6}$ eV/V, $\alpha_V= 1.9 \times 10^{-6}$ eV/V$^2$, Yellow: $d_V = -3.0 \times 10^{-6}$ eV/V, $\alpha_V =  1.8 \times 10^{-6}$ eV/V$^2$). Error bars are confidence intervals extracted from fits to the reflection line shape. Gray dashed lines indicate the voltage at which maximum energy occurs for the two parabolas, the gray arrow indicates the difference between them ($\Delta V$). \textbf{b.} Using data such as \textbf{a} at different $B_z$, difference between the extracted voltage ($\Delta V$) at which maximum energy occurs for $k_y^{-}$ and $k_y^+$ excitations.  \textbf{c.}  Change in $\alpha_V$ (polarizability) with $B_z$ (filled squares, left axis, and gray curve). Change  of lower polariton resonant energy with $B_z$ at $V_G=0$ V (filled circles, right axis, and red curve) from the value at $B_z=$ 0 T. Yellow and blue data points are data for $k_y^+$ and $k_y^-$, respectively.  \textbf{d.} Difference between extracted $d_V$ values (electric dipole moment) for polaritons propagating with $k_y^+$ and $k_y^-$. For {\bf b} -- {\bf d}, error bars are the estimated standard deviations of the mean for three repetitions of the experiment, and the solid lines are results of a numerical calculation (see text and the Supplemental information).
}
\label{magneticFieldDependence}
\end{figure}

In stark contrast, for $B_z = 5$ T the energy shift of polaritons with $E_{{\rm ext}}$ displays a significant difference between $\tilde E_{{\rm ext}}$ for $k_y^+$ and $k_y^- $, as illustrated in Figure~\ref{magneticFieldDependence}a. Magnetic field dependence of the difference of $\tilde E_{{\rm ext}}$ for $k_y^{\pm}$, Figure~\ref{magneticFieldDependence}b, shows a behaviour that is well described by a linear increase with $B_z$, demonstrating $E_{{\rm eff}}$ due to the Lorentz force.

The shape of the parabolas also contain further information about the $B_z$ induced changes to the polaritons. With increasing $B_z$, the polarizability decreases and the polariton energy at $E_{{\rm ext}}=0$ increases, as illustrated in Figure~\ref{magneticFieldDependence}c. The polariton energy at $E_{{\rm ext}}=0$ and the polarizability can be extracted from a fit to a second order polynomial: ${\epsilon_0} - \vec d \cdot {\vec E_{{\rm{ext}}}} - \alpha E_{{\rm{ext}}}^2$. The first order coefficient of the polynomial in turn, yields the induced dipole moment of the polaritons (excitons). The difference of the first order coefficient for $\pm k_y$ excitation beams (twice the induced dipole moment) are shown in Figure~\ref{magneticFieldDependence}d: we observe that the induced dipole moment first increases  with $B_z$ and then starts to slowly decrease for $B_z \ge 4$~T. We expect that for the high magnetic field regime, the induced dipole moment decreases as $\left| \vec d ~\right| \sim 1/B_z$~\cite{paquet1985}. We find very good agreement with a theoretical model of the polaritons (shown as solid lines in Figure~\ref{magneticFieldDependence}) that allows us to identify the physical mechanism underlying the overall $B_z$ and $E_{{\rm ext}}$ dependence (see Supplementary Information for the detailed information about the model). The change in the lower polariton energy is due to an interplay between a diamagnetic blue shift of the exciton transition energy and a red-shift due to an increase in the cavity-exciton coupling strength~\cite{pietka2015}. Concurrently, the polarizability decreases with $B_z$, as would be expected in a classical picture as the size of the polarizable particle decreases. We emphasize that due to the change in the exciton energy and cavity-exciton coupling strength the exciton content of the polaritons changes as $B_z$ is varied.  We also find that the energy shift of polaritons with $E_{{\rm ext}}$ is influenced by the decrease in the electron hole overlap due to $E_{{\rm ext}}$ induced polarization, which leads to a reduction of the cavity-exciton coupling strength. In the $E_{{\rm ext}}$ range that we probe the decrease in the cavity-exciton coupling leads to an effective smaller polarizability for polaritons as compared to excitons, due to a reduction on the excitonic character of the polaritons. Moreover, the model also captures the presence of $E_{{\rm eff}}$ due to the center of mass motion of polaritons and confirms that even with changes in exciton content with $B_z$, its expected dependence on $B_z$ is still linear. Finally, we emphasize that while $E_{{\rm eff}}$ arises from exciton physics, the strong coupling between excitons and photons has to be taken into account to obtain a quantitative agreement between the model and our experiments.

Remarkably, our findings demonstrate that polariton energy under perpendicular $B_z$ and $E_{{\rm ext}}$  depends not only on the magnitude of $k_y$ but also on its direction: this non-reciprocal flow of light is an indication of the presence of a gauge field for polaritons. The dispersion relation for free excitons propagating in the QW plane with wavevector $k_y \hat{y}$ in the presence of an electric field $\vec{E} = (\frac{\hbar}{M}k_y B_z + E_{{\rm ext}}) \hat{x}$ is:
\begin{equation}
{\epsilon _{{\rm{exc}}}}\left( {{k_y}} \right) = \epsilon_{\mathrm{exc}}(0) +\frac{\hbar^2 k_y^2}{2 M} - \alpha \left(\frac{\hbar}{M}k_y B_z + E_{{\rm ext}} \right)^2 =
\epsilon_{\mathrm{exc}}' + \frac{(\hbar k_y - q A_{{\rm eff}})^2}{2 M^{'}},
\end{equation}
where $M$ is the total exciton mass, and $\epsilon_{\mathrm{exc}}(0)$ is the exciton energy at $k_y=0$ and $E_{{\rm ext}}=0$.  $q A_{{\rm eff}} =  \frac{M'}{M} 2 \alpha B_z E_{{\rm ext}} $ is an effective vector potential for excitons, with $\epsilon_{\mathrm{exc}}'= \epsilon_{\mathrm{exc}}(0)-\alpha E_{{\rm ext}}^2-\frac{\left (q A_{{\rm eff}} \right)^2}
{2 M^{'}}$ and $M^{'} = M\left(1-\frac{\alpha B_z^2}{2 M}
\right)^{-1}$. The strong coupling to the cavity further changes the dispersion relation, but,  polaritons in a narrow energy-momentum range can still be treated as free particles under the influence of $q A$, whose value depends on the detuning between the excitons and polaritons as well as their effective masses (see Supplementary Information).

For our experiments both $B_z$ and $E_{{\rm ext}}$ are uniform over the area of our experiment yielding a constant gauge potential $qA$. Since physical observables are gauge invariant, we might expect our measurements to be independent of $qA$. In practice, our experiments involve propagation of photons between two regions with different constant gauge potentials; when calculating the magnitude of $q A$ above, we fixed a gauge by choosing the constant gauge potential for photons outside the cavity to be $qA = 0$. Conversely, $qA$ is observable in our experiments due to the presence of an effective (sheet) magnetic field at the interface between the cavity and vacuum that imparts kicks to the canonical momentum as photons/polaritons pass through the interface~\cite{Fang2013}.

To demonstrate the validity of the description of photon/polariton dynamics as determined by a gauge potential, we perform an interference experiment where we measure the phase accumulated by polaritons propagating for a length $l$ inside the sample relative to a fixed phase reference (see Figure~\ref{interferenceFigure}a and Methods for detailed information about the interference experiment). An interference image obtained at $E_{{\rm ext}}=0$ is shown in Figure ~\ref{interferenceFigure}b. Changes in the interference pattern with $y$ at two different $E_{{\rm ext}}$ are illustrated in Figure~\ref{interferenceFigure}c.

A plot of the phase change of polaritons with $E_{{\rm ext}}$ for different direction of excitation at $B_z = 6$ T is shown in Figure~\ref{interferenceFigure}d. The data shows that at a fixed $E_{{\rm ext}}$ and at the same excitation frequency, the phase accumulated for polaritons propagating along $\pm \hat{y}$ directions differ, and that this phase difference depends on the sign of $E_{{\rm ext}}$ and $B_z$. The in-plane momentum of the polaritons with a fixed energy is modified by the absence or presence of a constant vector potential so that the additional accumulated phase for traveling a distance $l$ is given by $\frac{q}{\hbar} \int \vec{A} \cdot dl$. At a fixed $E_{{\rm ext}}$ the difference between the phase accumulated when exciting with $k_y^{\pm}$ allows us to avoid phase changes due to the Stark effect and to directly obtain $\Delta \phi_{k_{y}^+} -\Delta \phi_{k_{y}^-} = 2 \frac{q}{\hbar} A l$. The fact that $\Delta \phi_{k_{y}^+} -\Delta \phi_{k_{y}^-}$ depends both on the sign of the electric field as well as the magnetic field is another manifestation of the validity of the description with $A^{{\rm eff}}$. Since the phase shift as a function of $E_{{\rm ext}}$ can be modeled with a second order polynomial in $E_{{\rm ext}}$, the dc-Stark effect is the dominant underlying mechanism for the phase shift. Figure~\ref{interferenceFigure}d shows that polaritons acquire a phase shift of $\sim 0.25$~radians  due to $qA$ as they propagate an average distance $l = 9\, \mu$m. Hence, the parameters of our experiment already ensure significant phase accumulation over the lattice constant ($\ge 2.4 \, \mu$m) of state of the art polariton lattices~\cite{Jacqmin2014} or average seperation ($ \sim 4\, \mu$m) of coupled polariton microcavities~\cite{Rodriguez2016}. Choosing detunings that yield predominantly exciton-like lower polaritons will further increase $qA$. In addition,  employing structures with longer polariton lifetimes~\cite{Steger2015} will allow the realization of  polariton lattices with larger lattice constants. With these advances, it should be possible to design structures where accumulated polariton phase due to $qA$ in a unit cell could be on the order of $\pi$.

Together with strong photon-photon interactions, realization of tunable artificial gauge fields is key for ongoing research aimed at observation of topological order in driven-dissipative photonic  systems~\cite{Hafezi2013,Umucalilar2012}. Cavity-polaritons constitute particularly promising candidates in this endeavour since their partly exciton nature enhances  interactions~\cite{Amo2009, Ferrier2010, Ferrier2011} and ensures time-reversal-symmetry breaking by external magnetic fields~\cite{Nalitov2015,Karzig2015}. Our work demonstrates that the same polariton system also allows for realization of tunable gauge fields, if an in-plane electric-field gradient is introduced in the QW. While such gradients are manifest in most gated structures, creation of a constant artificial magnetic field over a region of several $\mu$m should be possible in specially designed structures. We envision that lattices of dipolaritons~\cite{Cristofolini2012} would provide a very promising avenue towards this endeavour: the use of an in-plane magnetic field in this case would effect a gauge potential that is substantially enhanced by the large  dipolariton-polarizability and strong external electric fields along the growth direction~\cite{butov2001, lozovik2002}. The gradients of the latter in turn could be used to generate fluxes approaching magnetic flux quantum in a plaquette of area $\sim 5\times5 \mu m^2$. (Di)Polaritons subject to complex electric field distributions enabling large, tunable and inhomogeneous effective gauge field profiles~\cite{Imamoglu1996} would constitute a novel feature in exploration of topologically non-trivial non-equilibrium quantum systems.

\section{Methods}

\begin{figure}
\includegraphics[width = \textwidth]{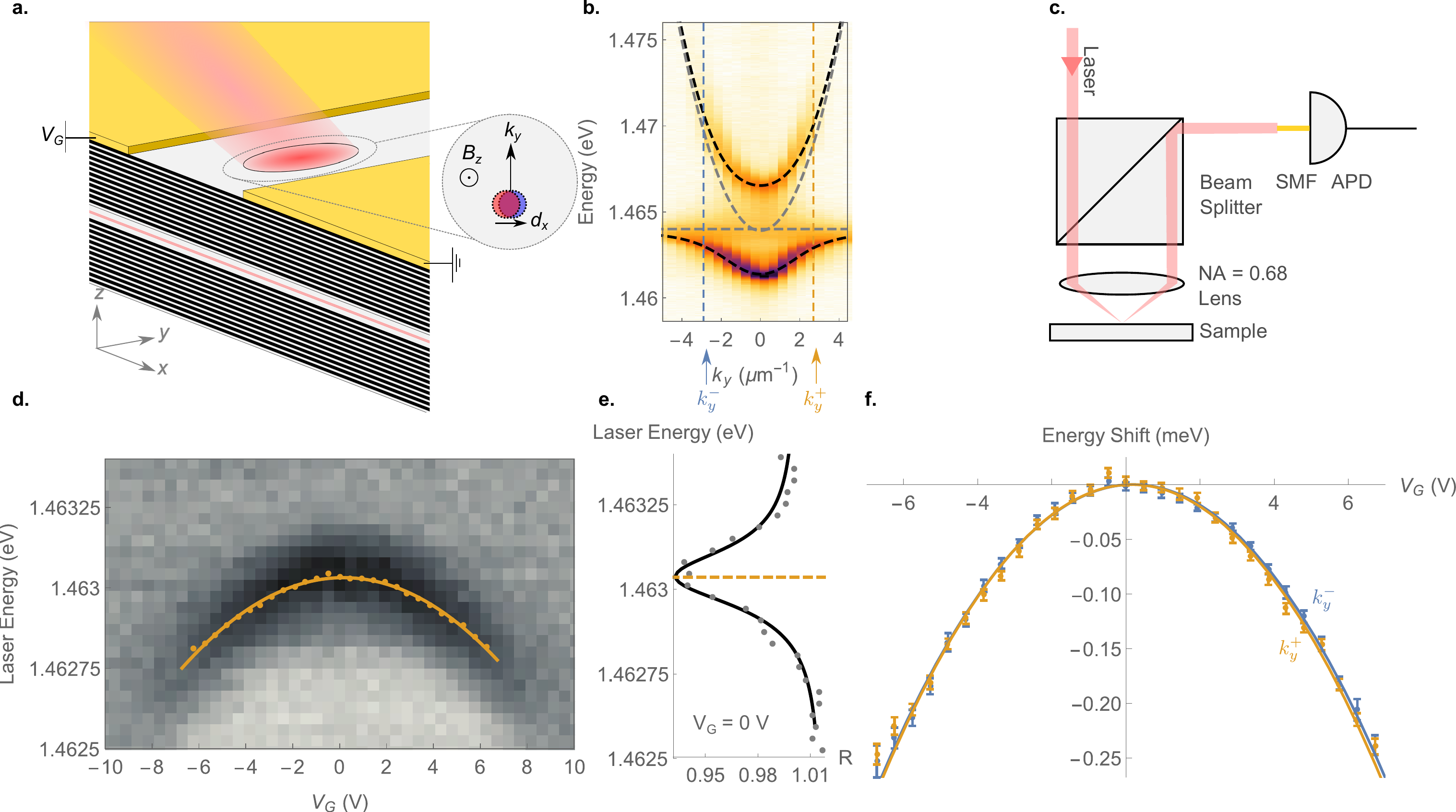}
\caption{\textbf{Description of the sample and characterization experiments at $B_z$ = 0 T.} \textbf{a.} The sample, held at 4 K in a helium bath cryostat, contains three 9.6 nm-thick In$_{0.04}$Ga$_{0.96}$As QWs located at an antinode of a cavity formed by two distributed Bragg reflectors (DBR). On the sample surface two metal gates are deposited with a 30 $\mu$m gap (see Supplementary Information). An electric potential $V_G$ applied to the gates creates an electric field in the $x$ direction. Inset: With a magnetic field in the $z$ direction polaritons excited with in plane wavevector $k_y$ exhibit a dipole moment $d_x$. \textbf{b.} $k_y$ resolved photoluminescence spectra at $B_z = 0$ T. The position on the sample is chosen such that for $|\vec{k}| \sim 0$ the detuning of the cavity mode energy from the exciton resonance is less than 0.09 meV  (Figure~\ref{ideaFigure}b) which is small compared to the exciton cavity coupling strength 5.2 meV. Gray dashed lines show the extracted bare cavity and exciton dispersions, black lines show polariton dispersions (see Supplementary Information for parameters), orange and blue lines correspond to $k_y = k_y^+$ and $k_y = k_y^-$, respectively. \textbf{c.} Experimental setup used in reflection measurements. Reflection of a linearly polarized laser beam ($<$ 100 pW) with in-plane wavevector $k_y$ is coupled to a single mode fiber (SMF) and its intensity is detected with an avalanche-photodiode (APD).  \textbf{d.} Changes in the reflected intensity of a laser beam exciting polaritons with in-plane wavevector $k_y^+$ as a function of laser energy and $V_G$. Yellow dots indicate the extracted polariton resonance energies. Note that the polariton signal is missing when large $V_G$ is applied ($V_G>$ 8 V) due to the ionization of the exciton. \textbf{e.} Line cut of the reflection spectrum at $V_G=0$ V. Gray points are the measured reflection data, black solid line shows the fitted lineshape (see Supplementary Information). Dashed yellow line indicates the extracted polariton resonance energy. \textbf{f.} Change of polariton energy from $V_G= 0$ V, as a function of $V_G$ for polaritons excited with  $k_y^+$ (yellow) and $k_y^-$ (blue) at $B_z=$ 0 T, solid lines are fits to second order polynomials (Blue: $1.9 \times 10^{-6}~V_G-5.9 \times 10^{-6} ~V_G^2$, Yellow: $2.2 \times 10^{-6} ~V_G-5.8 \times 10^{-6} ~V_G^2$). Error bars are confidence intervals extracted from fits.
} 
\label{ideaFigure}
\end{figure}

\textbf{In-plane electric field.} 
As described in Figure~\ref{ideaFigure}a,  we apply an electric potential $V_G$ between two gates to create $E_{\mathrm{ext}}$. We find a good agreement between our data and  a numerical calculation for the polariton behavior (see Supplementary Information) with the electric field $E_{\mathrm{ext}} = 0.84~V_G/(30\, \mu \mathrm{m})$. Since the two quantities are linearly related we use $E_{\mathrm{ext}}$ and $V_G$ interchangeably. For all experiments that involve a non-zero in-plane electric field, we acquire data in a pulse sequence (see Supplementary Information). This sequence is applied in order to avoid charge accumulation induced variation in the actual electric field.

\begin{figure}
\includegraphics[width = \textwidth]{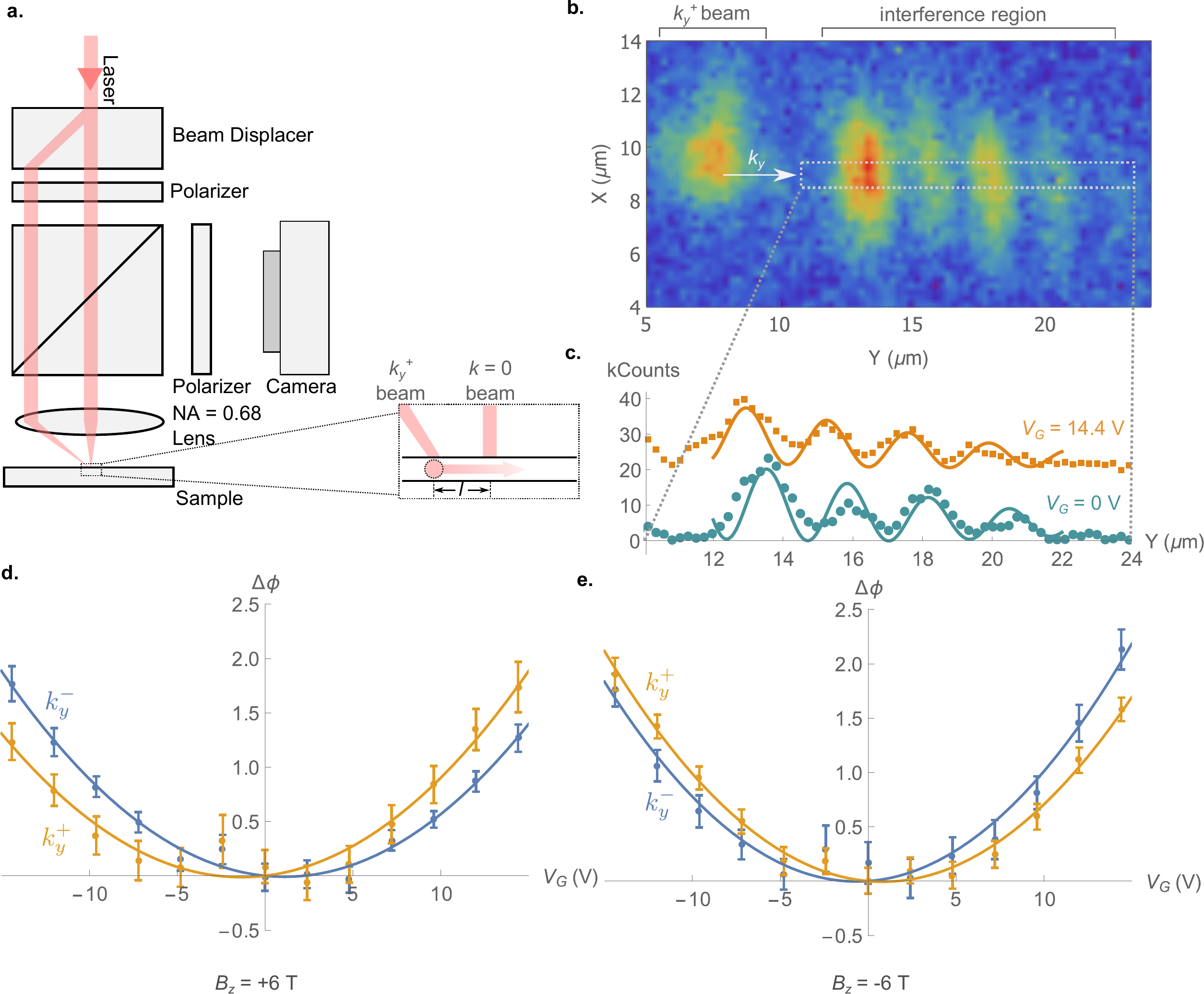}
\caption{\textbf{Demonstration of effective gauge potential for propagating polaritons.} \textbf{a.}
A single laser beam is split into two using a birefringent beam displacer, ensuring
the relative phase between the two beams do not fluctuate during the experiment. 
Resulting two linearly polarized beams (total power 40 nW) are 
incident at different positions on the high numerical aperture (NA = 0.68) lens.
One beam that is 1.2 mm off from the center of the lens is incident on the sample with $k_y^{\pm}$.
The second beam passes through the center of the lens and thus has vanishing in-plane momentum
($k = 0$ beam). The sample--lens distance is less than the focal length of the lens to ensure that the two beams
are incident on the sample with their centers displaced by $l\sim 9\, \mu$m. To detect polariton flow we use an additional linear polarizer between the camera and sample that transmits light that is nearly orthogonally polarized to the incident beams.  \textbf{b}. An interference image obtained at $B_z = - 6 $ T, $V_G = 0$ V. Polaritons excited by the $k_y^+$ beam at 1.46467 eV propagate in the direction indicated by the white arrow. In the spatial region where photons emitted by propagating polaritons overlap with the reflected $k = 0$ beam an interference pattern is observed. \textbf{c.} In green, sum of the detected intensity of the interference pattern shown in {\bf b} in the range $x = [8.5-9.4] \, \mu$m.  In yellow, same as for green but shifted in intensity, for an image obtained at $V_G = 14.4 $ V. Solid lines are fits to the model described in Methods and show a difference in the extracted phase.  \textbf{d.} Change in the extracted phase at $B_z = 6$ T as a function of $V_G$ for polaritons excited with $k_y^+ $ (yellow) and  $k_y^- $ (blue). Solid lines show second order polynomial fits to the phase change as a function of $V_G$ (Blue: $-0.016~V_G+0.0072~V_G^2$, Yellow: $0.019~V_G + 0.0071~V_G^2$). Error bars on the experimental data are based on the uncertainty extracted from the fit of the phase. \textbf{e.} Same as \textbf{d.} at $B_z = -6$ T, Blue: $0.011~V_G+0.0089~V_G^2$, Yellow: $-0.014~V_G + 0.0084~V_G^2$.
} 
\label{interferenceFigure}
\end{figure}

\textbf{Polariton interference measurement.} 
We use the setup depicted in Figure~\ref{interferenceFigure}a to create two  beams that are incident with different in plane momenta ($k_y^\pm$ and $k = 0$) and are separated by $l$.  The laser energy is chosen to be nearly resonant with the polariton modes with $k_y^{\pm}$. Thanks to the steep polariton dispersion, at the same energy, $k = 0$ beam is off resonance from the polariton modes;  the photons in this beam are therefore reflected by the top mirror without being subject to $A^{{\rm eff}}$. Photons in the $k_y^{\pm}$ beam are converted into a propagating polariton cloud. As polaritons propagate towards the position of the $k = 0$ beam, they accrue an additional phase due to $A^{{\rm eff}}$. Propagating polaritons continue to emit photons out of the sample, and when these photons overlap with those from the $k = 0$ beam they interfere, allowing us to measure changes in the accumulated phase. 

We model the light intensity in the region where the interference occurs (interference region) as the sum of a Gaussian ($k = 0$ beam) and a plane wave (polaritons propagating with $k_y$):
\begin{equation}
|b_0 e^{-(y-y_0)^2/\sigma_y^2} + a_0 e^{i \phi} e^{-i k_y y}|^2 + b_1. \nonumber
\end{equation}
We independently determine $b_0$, $\sigma_y$, and $y_0$ using images obtained at very high applied electric field for which the polariton term is negligible ($a_0 \ll b_0$), and fix $k_y$. For each $E_{{\rm ext}}$ we fit for $a_0$, $b_1$, and $\phi$ values.

Since our excitation beams have a finite size, they have a finite width in $k_y$ (0.45 $\mu$m$^{-1}$), therefore the exact $k_y$ value of the polaritons that are excited is determined by the energy of the excitation beam.  For a fixed excitation laser energy, changes in $E_{{\rm ext}}$ will shift the dispersion relation so that $k_y+\Delta k_y$ is resonant with the excitation laser. These changes (due to $A^{\mathrm{eff}}$ or Stark effect) in the dispersion relation show up as phase changes in our experiments, $\Delta \phi = l \Delta k_y$ where $l$ is the mean distance from the excitation spot to the interference region. Since we are interested in how the phase changes with $E_{\mathrm{ext}}$ and the absolute phase between the two beams is not determined, we set the phase at $E_{{\rm ext}}=0$ to 0 radian.

%\bibliographystyle{naturemag}
%\bibliography{export}

 \paragraph*{\textbf{Acknowledgements}}
The Authors acknowledge many insightful discussions with Iacopo Carusotto. This work is supported by SIQURO, NCCR QSIT and an ERC Advanced investigator grant (POLTDES).

\clearpage

\clearpage

\clearpage

\end{document}